
\input harvmac

\overfullrule=0pt
\parskip=0pt plus 1pt

\def\CC{\hbox{\bf C}}
\def\ZZ{\hbox{\bf Z}}

\def\l{\langle}
\def\r{\rangle}
\def\res{\hbox{Res}}

\def\mod#1{\overline{\cal M}_{0,#1}}


\def\np{Nucl. Phys.}
\def\pl{Phys. Lett.}

\def\cmp{ Comm. Math. Phys.}

\def\GINSMOORE{P. Ginsparg and G. Moore, {\it Lectures on 2-D Gravity
and 2-D String Theory}, Los Alamos and Yale Preprint LA-UR-92-3479,
YCTP-P23-92; hep-th/9304011.}
\def\KKLI{K. Li, \np\ {\bf B354} (1991) 711,725.}
\def\CECVAF{S. Cecotti and C. Vafa, as quoted in Ref.\mukhivafa}

\def\KS{Y. Kazama and H. Suzuki, \np\ {\bf B321} (1989) 232.}
\def\MUKHIVAFA{S. Mukhi and C. Vafa, \np\ {\bf B407} (1993) 667.}
\def\DVVLG{R. Dijkgraaf, H. Verlinde and E. Verlinde, \np\ {\bf B352}
(1991) 59.}
\def\VAFALG{C. Vafa, Mod. Phys. Lett. {\bf A6} (1991) 337.}
\def\DOUGLAS{D. Gross and A.A. Migdal, Nucl. Phys. {\bf B340} (1990)
333\semi
M. Douglas, \pl\ {\bf 238B} (1990) 176.}
\def\WITTENPHASE{E. Witten, \np\ {\bf B340} (1990) 281.}
\def\WITTENMATRIX{E. Witten, \np\ {\bf B371} (1992) 191.}
\def\LOSEV{A. Losev, ITEP Preprint PRINT-92-0563 (Jan
1993), hep-th/9211089.}
\def\EGUCHIETAL{T. Eguchi, H. Kanno, Y. Yamada and S.-K. Yang, \pl\
{\bf 298B} (1993) 73.}
\def\LOSPOLY{A. Losev and I. Polyubin, ITEP Preprint PRINT-93-0412
(May 1993), hep-th/9305079.}
\def\MATMOD{G. Moore, \np\ {\bf 368} (1992) 557\semi
G. Mandal, A. Sengupta and S. Wadia, Mod. Phys. Lett. {\bf A6}
(1991) 1465\semi
K. Demeterfi, A. Jevicki and A. Rodrigues, \np\ {\bf B362}
(1991) 173\semi
J. Polchinski, \np\ {\bf B362} (1991) 125.}
\def\DIFK{P. di Francesco and D. Kutasov, \pl\ {\bf 261B} (1991) 385.}
\def\COHOMOLOGY{
B. Lian and G. Zuckerman, \pl\ {\bf 266B} (1991) 21\semi
S. Mukherji, S. Mukhi and A. Sen, \pl\ {\bf 266B} (1991) 337\semi
P. Bouwknegt, J. McCarthy and C. Pilch, \cmp\ {\bf 145} (1992) 541.}
\def\WITZWIE{E. Witten and B. Zwiebach, \np\ {\bf B377} (1992) 55.}
\def\CY{B. Greene, C. Vafa and N. Warner, \np\ {\bf B324} (1989) 371\semi
E. Witten, \np\ {\bf B403} (1993) 159.}
\def\DIJKMP{R. Dijkgraaf, G. Moore and R. Plesser, \np\ {\bf B394}
(1993) 356.}
\def\KLEBLOWE{I. Klebanov and D. Lowe, \np\ {\bf B363} (1991) 543.}
\def\PENNER{J. Harer and D. Zagier, Invent. Math. 185 (1986) 457\semi
R.C. Penner, Commun. Math. Phys. 113 (1987) 299, J. Diff. Geom. 27
(1988) 35\semi
J. Distler and C. Vafa, Mod. Phys. Lett. A6 (1991) 259.}

{\nopagenumbers
\font\bigrm=cmbx10 scaled\magstep1
\rightline{MRI-PHY/13/93}
\rightline{TIFR/TH/93-62}
\rightline{hep-th/9312189}
\rightline{December 1993}
\ \bigskip\bigskip
\centerline{\bigrm TOPOLOGICAL LANDAU-GINZBURG MODEL}
\smallskip
\centerline{\bigrm OF TWO-DIMENSIONAL STRING THEORY}
\vskip 1truecm
\centerline{Debashis Ghoshal\foot{E-mail: ghoshal@mri.ernet.in}}
\vskip 5pt
\centerline{\it Mehta Research Institute}
\centerline{\it of Mathematics \&\ Mathematical Physics}
\centerline{\it 10 Kasturba Gandhi Marg, Allahabad 211 002, India}
\vskip 8pt
\centerline{Sunil Mukhi\foot{E-mail: mukhi@theory.tifr.res.in}}
\vskip 5pt
\centerline{\it Tata Institute of Fundamental Research}
\centerline{\it Homi Bhabha Road, Bombay 400 005, India}

\bigskip\bigskip
\centerline{ABSTRACT}

We study a topological Landau-Ginzburg model with superpotential
$W(X)=X^{-1}$. This is argued to be equivalent to $c=1$ string theory
compactified at the self-dual radius. We compute the tree-level
correlation function of $N$ tachyons in this theory and show their
agreement with matrix-model results. We also discuss the nature of
contact terms, the perturbed superpotential and the flow of operators
in the small phase space. The role of gravitational descendants in
this theory is examined, and the tachyon two-point function in genus 1
is obtained using a conjectured modification of the gravitational
recursion relations.
\ \vfill

\eject     }

\ftno=0

\beginsection 1. Introduction

Non-critical string theory in the background of matter with
central charge $c\le 1$ has been studied intensively over the
past few years. The case $c=1$ is the most interesting as there
is a propagating degree of freedom, the massless `tachyon', in
addition to the discrete states.  This theory also has a natural
physical interpretation of a critical string moving in a two
dimensional target space. For a comprehensive review, and
references to original papers, see Ref.\ref\ginsmoore{\GINSMOORE}.

Discretization of the world-sheet via random matrix models first
demonstrated the exact solvability of these models. This was
subsequently better understood in terms of a topological field
theory description. It was shown\ref\wittenphase{\WITTENPHASE}
that perturbations of pure topological gravity can reproduce an
infinite subclass of $c<1$ non-critical string models. The
remaining models in the $c<1$ series are obtained by coupling
specific topological matter systems to topological
gravity\ref\kkli{\KKLI}. The relevant matter theories have a
discrete series of topological central charges $\hat{c} =
k/(k+2)$ where $k$ is a positive integer..

The problem of finding a topological description of the $c=1$
string remained unsolved for a considerable amount of time. In
particular, naive attempts to continue the topological series
for $c<1$ by taking the limit of large $k$ do not seem to work.
However, more recently it was
understood\ref\mukhivafa{\MUKHIVAFA} that the correct
topological field theory is the twisted
Kazama-Suzuki\ref\ks{\KS} coset model SL$(2)_k/$U$(1)$ with
$k=3$.  Another way to view this coset theory is to think of it
as an SU$(2)_k/$U$(1)$ coset, where the level $k$ is continued
to the value $-3$.  In this theory the topological central charge
is $\hat{c}=3$, the `critical' value for topological field
theories\wittenphase.

This model has been argued to reproduce $c=1$ string theory where the
$c=1$ scalar field is compactified at the self-dual radius. It is
shown in Ref.\mukhivafa\ that the entire spectrum of physical states
in the latter theory (at zero cosmological constant) can be reproduced
through a double cohomology in the coset model\foot{In this picture,
the cosmological operator is a screening-charge-like perturbation of
the CFT, just as in the Liouville formalism, while one works in the
Hilbert space of the unperturbed theory.}. More important, a
Lagrangian description of the same coset
model\ref\wittenmatrix{\WITTENMATRIX} (which was previously used for
the SU$(2)_k/$U$(1)$ case with positive $k$), gives explicit results
which can be continued to $k=-3$. This remarkably leads to the
evaluation of the genus-$g$ partition function and certain tree-level
4-point functions, which agree with the corresponding results in the
$c=1$ matrix model (at self-dual radius) with {\it nonzero}
cosmological constant, after a natural identification of chiral
primaries with tachyons. Although this formulation in principle
determines all correlators in every genus, it is not so easy to
extract explicit expressions for correlators other than those
mentioned above.

This motivates us to turn to the Landau-Ginzburg (LG) description of
topological matter, which has proved to be most convenient for
explicit calculation in the context of minimal matter
backgrounds\ref\dvvlg{\DVVLG}\ref\vafalg{\VAFALG}. In this description
the LG superpotential determines the properties of the theory. The
SU$(2)_k/$U$(1)$ coset theory for positive integer level $k$ is
described by one chiral superfield X with the superpotential
$W(X)={1\over k+2}X^{k+2}$. The physical operators defined by the BRS
cohomology of the topological string correspond to the chiral
primaries, while the topological algebra given by the factorization
(of the zero-form operators) is derived from the isomorphism to the
chiral ring $\CC[X]/W'(X)$. In Ref.\dvvlg, tree-level correlators of
the topological field theory were calculated and shown to agree with
the results obtained from matrix models. Moreover, a direct
correspondence between the LG superpotential and the KdV differential
operator\ref\douglas{\DOUGLAS}\ of matrix models was demonstrated.

Since the topological coset SU$(2)_k/$U$(1)$ and the
Landau-Ginzburg models with superpotential $W(X)={1\over
k+2}X^{k+2}$ give equivalent descriptions of $c<1$ matter
coupled to gravity, it is reasonable to guess that the
correspondence holds even when the level $k$ is continued to $-3$.
Thus we are led to consider the LG theory with superpotential
$W(X)=-X^{-1}$, as a candidate for a topological description of
compactified $c=1$ string theory. (It has already been
observed\ref\cecvaf{\CECVAF} that the tachyon 4-point functions
of $c=1$ string theory are obtainable from the LG approach).

This theory requires extra work to define it precisely, as the
superpotential is not polynomial and many of the properties described
above, including the existence of a nilpotent chiral primary ring, do
not strictly hold. The difficulties involved here are analogous to
those encountered in defining an SL$(2)_k$ conformal field theory,
where (unlike the case of SU$(2)_k$) it is not possible to rigorously
determine the integrable representations and other basic properties
just by analysing the Lagrangian.  Nevertheless, consistency
requirements are a powerful constraint in the study of
Landau-Ginzburg theories, and we will see below that these suffice to
find reasonable spectra of physical operators and formulae for
correlation functions. We will show that the correlators of $c=1$
string theory follow in a very simple way from our considerations,
and moreover the complementary pictures of Refs.\dvvlg\ and
\ref\losev{\LOSEV} both have natural analogues in the present model.

In the present work we will examine this theory in some detail and
extract several of its properties which not only confirm its
identification with the $c=1$ string (directly at nonzero cosmological
constant), but also reveal some interesting and unexpected
features. We will then identify the gravitational descendants, and
make a hypothesis for the gravitational recursion relations in this
theory. With this hypothesis we show that one can explicitly obtain
correlators on the torus which agree with matrix-model results. We
also study the flow of the model in coupling-constant space.

\beginsection 2. Landau-Ginzburg Tachyons and Selection Rules.

Correlators of local operators in conventional LG theory (with
superpotential ${1\over k+2}X^{k+2}$) are obtained from the knowledge
of the three-point function (structure constant) of three primaries on
the sphere, $c_{ijl}$, and the two-point function (metric)
$\eta_{ij}$. To compute correlators involving integrals of two-form
operators, one computes the perturbed ring corresponding to the
perturbed superpotential. The perturbations are by the physical
operators, the chiral primaries $\phi_i$.  Let $t_i$, $0\le i\le k$,
denote the coupling constants corresponding to the scaling fields
$\phi_i(t)$ of the perturbed ring. The central object to calculate is
the perturbed structure constant $c_{ijl}(t)$---differentiating this
an appropriate number of times and setting $t_i=0$ for all $i$, gives
all the correlators of the original theory. In this picture, the
superpotential $W$ and the primaries evolve as a function of $t$, that
is they {\it flow} in the small phase space, whose coordinates are the
couplings $t_i$.

After coupling this theory to gravity there arise an infinite number
of physical operators, the so called gravitational descendants of the
matter primaries. Correlators involving descendants can, however, be
reduced to those of the primaries alone, by the use of the recursion
relations\wittenphase\kkli.  Recently, Losev has shown that the LG
description is adequate for the topological theory even after its
coupling to gravity\losev\ (see also \ref\eguchietal{\EGUCHIETAL}).
The $m$-th descendant of the primary $X^i$, $0\le i\le k$, in this
description, can be written as $\sigma_m(X^i)\sim X^{i+m(k+2)}$.
These are matter secondaries which do not decouple in topological
gravity.  In Ref.\ref\lospoly{\LOSPOLY}, a recursive prescription is
given to calculate the tree-level $N$-point correlator by reducing it
in terms of $(N-1)$-point correlators.  (It is important that the
$N$-th insertion is a primary.)

With this background, let us turn now to the study of the LG
model with superpotential $W(X)=-X^{-1}$. The superpotential
must have U$(1)$ charge 1, which implies that the field $X$ has
U$(1)$ charge $-1$ and the various powers $X^i$ have charge $-i$.
We start by assuming that the physical fields are all (positive
and negative) powers of X. It is not yet clear whether these
should all be treated as gravitational primaries, and the
distinction between primaries and secondaries in this set will
become clear below.

Let us now examine the selection rules based on U$(1)$ charge.  It is
known for LG theories of topological central charge ${\hat c}$ coupled
to gravity, that the correlators satisfy anomalous U$(1)$ charge
conservation laws\kkli\dvvlg\wittenmatrix. Suppose we consider the
genus-$g$ $N$-point function of gravitational primaries, each carrying
U$(1)$ charge $q_i$, then we have the selection rule
\eqn\anomcharge{\sum_{i=1}^N (q_i - 1) = (g-1)(3-{\hat c})}
Note that this selection rule is also true for gravitational
secondaries, if we assign an integer effective U$(1)$ charge $m$ to
the $m$th gravitational secondary $\sigma_m$\kkli. This fact will be
useful in what follows.

Precisely at ${\hat c}=3$, which is the case of interest here,
we find the universal conservation law $\sum_{i=1}^N (q_i -1)$,
valid in every genus. Now in the $c=1$ string there is a unique
conserved quantity in every genus, and that is tachyon momentum.
Thus we are led to identify LG fields of U$(1)$ charge $k+1$ to
discrete tachyons of momentum $-k$ (we choose the last sign for
later convenience, as clearly a choice of convention is required
here. This is due to the $\ZZ_2$ symmetry of the $c=1$ string.
In principle there could be an arbitrary factor as well,
but that must equal 1 since both the tachyons of the self-dual
compactified $c=1$ string and the monomials of the LG theory are
labelled by integers). Thus we are led to the correspondence
\eqn\tachcorresp{T_k = X^{k-1}}
Later we will return to questions of overall normalization.

The zero-momentum tachyon $T_0=X^{-1}$ becomes the cosmological
operator in this identification. Note that this coincides with the
superpotential of the theory, which is one reason to anticipate that
the theory we are constructing is already at unit cosmological
constant. The cosmological operator satisfies U$(1)$ charge
conservation for any number of insertions and in any genus, as it
should. Note also that tachyons of positive and negative momenta
should be thought of as having opposite target-space chirality.

The three-point function determines the structure constants of this
theory. We can simply insert our identification of Landau-Ginzburg
tachyons into the standard formula, which gives
\eqn\struc{
\eqalign{c_{k_1 k_2 k_3} &= \l T_{k_1}T_{k_2}T_{k_3}\r \cr
&= \res\left({X^{k_1-1}X^{k_2-1}X^{k_3-1}\over X^{-2}}\right)\cr
&= \delta_{k_1+k_2+k_3,0}\cr}}
Here, `Res' means the residue obtained by contour-integrating the
argument in $X$ around the circle at infinity in the complex
$X$-plane.

{}From the structure constants we obtain the metric:
\eqn\metric{\eta_{k_1 k_2} = c_{k_1 k_2 1} = \delta_{k_1 + k_2+1,0}}
Note that for the metric, we must take the third index of the
structure constant to correspond to the puncture operator, which
actually carries a momentum $k=1$ (see Eq.\tachcorresp).

We see that it was really necessary to take both positive and
negative powers of the LG superfield $X$, since they are dual to each
other. In a certain sense, this suggests that all the monomials in
$X$ and $X^{-1}$ should be thought of as gravitational primaries.
However, this interpretation cannot be taken literally as it will
lead to incorrect results in computing $N$-point functions, as we
will see in the next section. It will turn out that in
correlation functions, only the positive-momentum tachyons
behave as primaries, while the negative-momentum tachyons behave
as gravitational secondaries of the cosmological operator.

\beginsection 3. Contact Terms and Four-point Function

The four-point function is the first non-trivial correlator, as it
involves an integration over the moduli space $\mod 4$. There are two
standard ways to compute this in theories of topological matter
coupled to topological gravity. One is to evaluate the structure
constants in the presence of a perturbation\dvvlg:
\eqn\fourptpert{
c_{k_1 k_2 k_3}(t_4) = \res\left({T_{k_1}(t_4)T_{k_2}(t_4)
T_{k_3}(t_4)\over (W + t_4T_{k_4})'}\right)}
where on the right hand side one uses not only the perturbed
superpotential, but also the perturbed forms of the fields, to first
order in $t_4$. Differentiating this structure constant
in $t_4$ gives the genus-0 4-point function $\l
T_{k_1}T_{k_2}T_{k_3}T_{k_4} \r_W$. We will perform this calculation
below, but first we turn to the other computational procedure.

In the method of Ref.\losev, the four-point function is obtained by
differentiating the structure constants, which are however calculated
leaving the fields unperturbed. Then one adds in contact terms
arising from the collision of the fourth field with the other three.
The result is
\eqn\fourpt{\eqalign{
\l T_{k_1}T_{k_2}T_{k_3}T_{k_4}\r_W =& {\del\over\del t_4}
\l T_{k_1}T_{k_2}T_{k_3}\r_{W+t_4 T_{k_4}}\Big|_{t_4=0} +
\l C_W(T_{k_4},T_{k_1})T_{k_2}T_{k_3}\r_W \cr &+
\l T_{k_1}C_W(T_{k_4},T_{k_2})T_{k_3}\r_W
+ \l T_{k_1}T_{k_2}C_W(T_{k_4},T_{k_3})\r_W,\cr} }
where, $C_W(T_{k_i},T_{k_j})$ is the contact term between the fields
$T_{k_i}$ and $T_{k_j}$. The distinction between contributions from
the bulk and boundaries of moduli space is explicit in \fourpt.
The contact term is the contribution from the boundary,
where the positions of two fields collide.

The contact term for the LG theory in question is determined by
self-consistency. For LG theories with $k>0$, it was shown in \losev\
that the contact term between two fields $\phi_i,\phi_j$ is given by
$C_W(\phi_i,\phi_j) = {d\over dX}\left({\phi_i(X)\phi_j(X)\over
W'(X)}\right)_+$, where the subscript denotes the prescription of
keeping only the positive powers of $X$ in the resulting expression.
This is the unique choice leading to a symmetric expression for the
four-point function. It is clear that this contact term gets
contributions only if the OPE of the two fields which collide gives
rise to a gravitational secondary.

It turns out that the same contact term (with a change of sign of the
powers) is appropriate in the $k=-3$ model as well. We find that with
our conventions, we must take
\eqn\contact{
C_W(T_{k_i},T_{k_j}) = {d\over
dX}\left({T_{k_i}(X)T_{k_j}(X)\over W'(X)}\right)_-}
where the subscript indicates that we keep only the negative powers of
$X$. Symmetry of the four-point correlator follows from the
identity
\eqn\identity{
\res\left\{{d\over dX}\left({P\over W'}\right)_-{Q\over W'}
- {P\over W'}{d\over dX}\left({Q\over W'}\right)_-\right\} =
{1\over 2}\res\left\{{P'Q - PQ'\over (W')^2}\right\}}
for two polynomials $P,Q$ in $X$ and $X^{-1}$. Notice that the
relevant identity for positive integer level $k$ has the subscript $+$
corresponding to keeping positive powers of $X$ in the LHS of
\identity.

The contact term \contact\ between tachyons $T_{k_i}$ and
$T_{k_j}$ is thus
\eqn\contacterm{
C_W(T_{k_i},T_{k_j}) = (k_i+k_j) T_{k_i+k_j} \theta(-k_i-k_j),}
where $\theta(x)$ is the step function.
Using Eq.\fourpt\ and the contact term \contacterm, it is
easy to show that the four-point function is
\eqn\fourfinal{
\eqalign{
\l T_{k_1}T_{k_2}T_{k_3}T_{k_4}\r &= \delta\;(\sum_{i=1}^4 k_i)\;
(1 - \hbox{max}|k_i|)\cr
&= \delta\;(\sum_{i=1}^4 k_i)\;
[-\half|k_1+k_2| - \half|k_1 + k_3| - \half|k_2 + k_3| +  1]\cr}}
This result agrees with the tree-level correlator of four tachyons
calculated in matrix models\ref\matmod{\MATMOD}, evaluated at $\mu =
-1$. The same correlator was also computed in ref.\ref\difk{\DIFK} in
the continuum method.

In fact, in Ref.\difk, a compact expression for $N$-point
function of tachyons on the sphere is given. The correlator in
question is in the kinematic region $k_2,\cdots,k_N >0$ and
$k_1=-(k_2+\cdots+k_N)<0$. The answer can be expressed as
\eqn\difksnpt{
\l T_{k_1}\cdots T_{k_N} \r = \left({\del\over\del\mu}\right)^{N-3}
\mu^{-k_1-1}}
In the following section we will rederive this result from the LG theory.

We are now in a position to make more precise the identification of
the LG tachyons $T_k$ with those in the standard continuum approach.
Since the result of Ref.\difk\ is for the tachyons of so-called plus
dressing\ref\cohomology{\COHOMOLOGY}\ref\witzwie{\WITZWIE}, it is
natural to identify the $T_k$ to the positively dressed tachyons. The
integer $k$, in units of $\sqrt 2$, does indeed give the momentum of
the continuum tachyon. Notice that in the continuum language, the LG
tachyons are already {\it renormalized}, that is they are related to
the tachyon fields with their leg-factors absorbed\foot{It is
of course the {\it renormalized} cosmological operator which
is actually the correct cosmological operator $\varphi e^{\sqrt
2\varphi}$ of the $c=1$ string theory.}.

Before closing this section, let us point out that our results
are automatically those for cosmological constant $\mu=-1$. As
mentioned above, this is presumably due to the fact that the
superpotential is just the cosmological operator itself. This
idea can be tested by replacing the superpotential $-X^{-1}$ by
$\mu X^{-1}$, in which case we should expect to obtain
correlators with the right $\mu$-dependence. Remarkably, this is
just what happens. This will be shown in the section on
gravitational descendants below.

\beginsection 4. Multipoint Correlators of Tachyons

The $N$-point function in LG theory is related to the
$(N-1)$-point function by the recursion formula of
Ref.\losev\lospoly
\eqn\recursion{
\eqalign{
\l T_{k_1}\cdots T_{k_N}\r_W =& {\del\over\del t_N}
\l T_{k_1}\cdots T_{k_{N-1}}\r_{W+t_N T_{k_N}}\Big|_{t_N=0}\cr
&+ \sum_{i=1}^{N-1}
\l T_{k_1}\cdots C_W(T_{k_N},T_{k_i})\cdots T_{k_{N-1}}\r_W.}}
Unlike the expression for the four-point function, however, \recursion\
is true only if the $N$-th insertion is a primary. We will see below
that one recovers the expected matrix-model results only if the
inserted fields are {\it positive-momentum} tachyons, $T_k$ with
$k\ge 0$. Thus we are led to think of only $k\ge 0$ tachyons as
gravitational primaries. Fortunately, we can conveniently identify
the $k<0$ tachyons as gravitational descendants (this would not have
been possible to do with the $k\ge 0$ tachyons, as we will see).
Eventually we will also check that $k<0$ tachyons obey
recursion relations for descendants.

Using Eqs.\fourpt,\contacterm\ and \recursion, one can systematically
calculate correlators of 5, 6,$\ldots$ tachyons (with the restriction
that $k_5,k_6,\ldots$ are all positive). In the kinematical region,
$k_2,\cdots k_N>0$ and $k_1=-\sum_{i=2}^N k_i$, we now show that the
answer turns out to be
\eqn\nptfn{
\l T_{k_1}\cdots T_{k_N}\r = \prod_{i=1}^{N-3} (k_1 + i) }
in perfect agreement with Eq.\difksnpt\ evaluated at $\mu=-1$!

We prove \nptfn\ in the general case by induction. Notice first that
the contribution from the contact terms in \recursion\ simplifies for
the chosen kinematics, since only the contact term between $T_{k_N}$
and $T_{k_1}$ is non-zero. Using Eqs.\contacterm\ and the induction
hypothesis \nptfn, this contribution is
\eqn\dmgive{
\sum_{i=1}^{N-1}
\l T_{k_1}\cdots C_W(T_{k_N},T_{k_i})\cdots T_{k_{N-1}}\r_W =
\prod_{i=1}^{N-3} (k_1+k_N+i-1). }
Notice that for $k_N=1$, this is already the correct answer. This is
consistent as in that case, the contribution from the first term
vanishes (since the potential gets perturbed by a constant, leading
to no change in the structure constants).

To find the contribution due to the first term, we will first prove the
identity
\eqn\lemone{\eqalign{
{\del\over\del t_N}\cdots{\del\over\del t_{m+1}}\l T_{k_1}\cdots T_{k_m}
\r_{\tilde W}\Big|_{t_a=0} =& \prod_{i=1}^{N-3}(k_1+i)
- \sum_{a=m+1}^N\prod_{i=1}^{N-3}(k_1+k_a+i-1)\cr
&- \sum_{{a<b\atop m+1}}^N\prod_{i=1}^{N-3}(k_1+k_a+k_b+i-2) +
\cdots\cr
&+(-)^{N-m+1}\prod_{i=1}^{N-3}(k_1+\!\sum_{a=m+1}^N\!
k_a+i-N+m),\cr} }
where $\tilde W = W+\sum_{a=m+1}^N t_a T_{k_a}$.  This
identity is obviously true for $m=3$: The RHS is a polynomial of
degree $(N-3)$ in $k_4,\cdots,k_N$ and has zeroes at each of
$k_i=1$, $i=4,\cdots,N$; hence it is proportional to
$\prod_{i=4}^N(k_i-1)$. The polynomial is uniquely determined by
the coefficient of the term $k_4\cdots k_N$, which is easily
found to be $(-)^{N-3}(N-3)!$. On the other hand, this is
exactly what we get by explicit differentiation of the LHS. We
will now assume that \lemone\ true for an arbitrary $m$ and
prove by induction.

The LHS of \lemone\ can be written as
\eqn\lhs{
{\del\over\del t_N}\cdots{\del\over\del t_{m+1}}\Bigg[{\del\over\del t_m}
\l T_{k_1}\cdots T_{k_{m-1}}\r_{\tilde W+t_m T_{k_m}}\Big|_{t_m=0} + \l
C_{\tilde W}(T_{k_m},T_{k_1})T_{k_2}\cdots T_{k_{m-1}}\r_{\tilde W}
\Bigg]_{t_a=0}. }
Now, we know the first term from the induction hypothesis \lemone, and
therefore to prove the above identity, we need to show that,
\eqn\lemtwo{\eqalign{
{\del\over\del t_N}\cdots{\del\over\del t_{m+1}}
\l C_{\tilde W}(T_{k_m},T_{k_1})T_{k_2}&\cdots T_{k_{m-1}}\r_{\tilde W}
\Bigg|_{t_a=0} = \;\prod_{i=1}^{N-3}(k_1+k_m+i-1)\; \cr
&-\sum_{a=m+1}^N\prod_{i=1}^{N-3}(k_1+k_m+k_a+i-2)
+ \cdots\cr
&+(-)^{N-m+2}
\prod_{i=1}^{N-3}(k_1+k_m+\!\sum_{a=m+1}^N\!k_a+i-N+m-1).
\cr} }
To prove Eq.\lemtwo\ above, note that we can once again use the
induction hypothesis as the correlator involved in \lemtwo\ is
an $(m-1)$-point function. Expanding the contact term $C_{\tilde
W}(T_{k_m},T_{k_1})$ order by order in $t$ and using the
induction hypothesis, it is tedious but straightforward to show
that \lemtwo\ is true. This completes the proof of
\lemone.

A special case of \lemone\ is $m=N-1$. In this case the LHS is precisely
the first term is \recursion. Substituting for this from \lemone, we get
the desired result \nptfn.

Finally, one can try to go through the same procedure as above but
using a negative-momentum tachyon as the perturbing field. Already at
the level of the 5-point function, this gives an answer which differs
from the correct one, which shows that one cannot consistently take
negative-momentum tachyons to be gravitational primaries in this
theory. We will discuss in a subsequent section how they can in fact
be thought of as gravitational descendants, but contructed entirely
in terms of matter-sector variables, analogous to the picture
advocated by Losev\losev\ for the positive-level LG theory.

Actually from Eq.\contact\ we already find support for this picture.
The contact term in this equation arises only if the OPE of the
colliding fields results in a negative-momentum tachyon. Identifying
these tachyons as secondaries leads to the conclusion that the contact
terms in our theory arise only when a secondary is produced, exactly
as in the $k>0$ theories (see the comment below Eq.\fourpt).

\beginsection 5. Flow in the Small Phase Space

So far, we have been working in the basis of fields of Ref.\losev\lospoly,
where the fields as well as the superpotential, (except in intermediate
stages of computation), are independent of the couplings $t$. There exists
another picture, that of Ref.\dvvlg, where they acquire non-trivial
$t$-dependence---that is, they are said to {\it flow}. The relation
between the two is through the formal generating function of the
$t$-dependent multipoint correlator $\l T_{k_1}\cdots T_{k_N}\r_W (t)
\equiv
\l T_{k_1}\cdots T_{k_N} e^{\sum_{k_i>0} t_i T_{k_i}}\r_W$. This is equal
to the multipoint correlator $\l T_{k_1}(t)\cdots T_{k_N}(t)\r_{W(t)}$ in
the $t$-dependent picture.

The $t$-dependent fields and superpotential are solutions to the
differential equations\losev\lospoly
\eqn\relation{\eqalign{
{\del\over\del t_i} T_{k_j}(t) &= C_{W(t)}(T_{k_i}(t),T_{k_j}(t)),\cr
{\del\over\del t_i} W(t) &= T_{k_i}(t),\cr}}
where, the index $i$ is restricted to tachyons with positive $k_i$
(primaries) only. It is easy to explicitly integrate Eqs.\relation. To
this end notice that the contact term \contacterm\ between two tachyons with
positive momenta vanishes. This leads to a great simplification. First of
all, the primaries do not flow at all,
\eqn\primflow{
T_k(t) = T_k \quad\hbox{for } k>0,}
while the superpotential is only a linear function of the $t_i$
\eqn\pertpot{
W(t) = - X^{-1} + \sum_{i=1}^\infty t_i X^{i-1}. }

The flow for the secondaries, that is tachyons with negative
momenta, can be determined order by order in $t$. After a little
algebra, the solution ($T_k(t)$ for $k<0$) is found to be
\eqn\seconflow{\eqalign{
T_k(t) =& X^{k-1} + \sum_{{k_i>0\atop k+k_i<0}} t_i (k+k_i) X^{k+k_i-1} +
\sum_{{k_i\ne k_j>0\atop k+k_i+k_j<0}}
t_i t_j (k+1)(k+k_i+k_j) X^{k+k_i+k_j-1}\cr
&+ \sum_{{k_i\ne k_j\ne k_l>0\atop k+k_i+k_j+k_l<0}}
t_i t_j t_l (k+1)(k+2)
(k+k_i+k_j+k_l) X^{k+k_i+k_j+k_l-1} + \cdots\cr} }
Each term on the RHS is totally symmetric in all the $t$'s, which is
just the statement that the flows in the different directions commute.

It is now easy to calculate the perturbed three-point function between
tachyons with momenta $k_1<0$ and $k_2,k_3>0$:
\eqn\threeptpert{
\eqalign{
c_{k_1 k_2 k_3}(t) =& \res\left({T_{k_1}(t) T_{k_2} T_{k_3}\over
W'(t)}\right) \cr
=& \delta_{k_1+k_2+k_3,0} + \sum_i t_i (k_1+1)\delta_{k_1 + k_2 + k_3
+ k_i,0}\cr
&+ \sum_{i,j} t_i t_j (k_1+1)(k_1+2)\delta_{k_1 +k_2 + k_3 +
k_i + k_j,0} + \cdots\cr}}
Differentiating this $(N-3)$ times and setting all the $t_i$ to 0, we
get the $N$-point correlator \nptfn, as expected.

For other choices of the kinematics, it is equally straightforward to
check that the correlators computed in the two bases give the same
answer. As an illustration, we have computed the 5-point function in
the configuration $k_1,k_2 < 0$ and $k_3,k_4,k_5 > 0$ in both the
approaches. There are several different kinematic regions within this
configuration. One example is the region $|k_2|>|k_5|>|k_1|>(|k_3|,
|k_4|)$, for which the 5-point function is
\eqn\fivept{\l T_{k_1} T_{k_2} T_{k_3} T_{k_4} T_{k_5} \r =
(k_2+1)(k_1 - k_5 +2)}

The fact that we get the correct correlators in the basis of
Ref.\dvvlg\ with the use of Eqs.\relation\ serves, in particular, as
an additional justification for our division of the Landau-Ginzburg
fields into primaries and secondaries according to whether they are
positive or negative momentum (and chirality) tachyons.

There is one important difference between our LG theory and those with
positive level $k$. In the latter, the two point function between
primaries is independent of the time $t$. The set of primary fields
defines the tangent space at the fixed point (conformal point) in the
space of topological theories. The two-point function defines a metric
in this space, and since it is independent of $t$, this metric is
constant and hence flat. On the other hand for the LG theory with
$k=-3$, the non-vanishing two-point function is between a primary and
a secondary, and from \seconflow, we see that it acquires a
non-trivial dependence on $t$.

\beginsection 6. Gravitational Descendants

The main point of Ref.\losev\ is that for $k>0$ LG matter coupled to
topological gravity, the gravitational descendants can be expressed
entirely in terms of matter degrees of freedom. Since in our case the
negative-chirality tachyons cannot be thought of as primaries, it is
reasonable to suppose that in fact they are gravitational secondaries,
expressed in terms of the matter sector fields. From U$(1)$ charge
conservation, we see that $T_k$ has the same charge as
$\sigma_m(T_{k'})$ whenever $k=k'-m$, in other words, taking the $m$th
gravitational secondary lowers the charge by $m$ units. Thus, assuming
that the only primaries are $T_k$ with $k>0$, it is possible to
construct secondaries of these which have the right charge to be
negative-chirality tachyons. In fact, this can be done in infinitely
many ways. We will see, however, that it is most natural to consider
the negative-chirality tachyons in a unique way as descendants of the
cosmological operator. Note that since gravitational descendants
effectively lower the momentum, it would not have been possible for us
to treat negative-chirality tachyons as primaries and
positive-chirality as secondaries, once our initial conventions are
fixed. Thus our hypothesis is consistent with the structure of the
theory in a nontrivial way.

According to \losev, the $m$-th descendant of the primary $\phi(X)$
can be written as
\eqn\desc{
\sigma_m(\phi) = W'(X)\int^{X}\!dX_m \; W'(X_m)\cdots
\int^{X_3}\!dX_2\; W'(X_2)\int^{X_2}\! dX_1 \;\phi(X_1). }
Our strategy is to construct descendants of the cosmological operator,
using the above prescription, and then compute a 4-point function in
two ways: once using purely matter variables, which we have already
done above, and then again using the gravitational descent
equations\wittenphase. Agreement between the two provides strong
evidence for our identification.

{}From Eq.\desc\ above, we have in our case
\eqn\secondary{
\sigma_m(T_0) = \sigma_m(X^{-1}) = X^{-2}\int^{X}\! dX_m\; X_m^{-2}
\cdots\int^{X_3}\! dX_2 \;X_2^{-2} \int^{X_2}\! dX_1 \; X_1^{-1}}
We see immediately that the very first integral produces a
logarithmic factor, and that this is present only in our LG theory
(which corresponds to the $c=1$ string) and not in the LG theories
which are associated to $c<1$ strings. This is evidently an
echo, in the topological description, of the logarithmic scaling
violations in the $c=1$ string. We will simply imagine that the
integral is `regularised', by slightly shifting the powers of $X$,
and at the end we will have a divergent $\Gamma$-function. This
will ultimately cancel out from both sides of our calculation.

With this understanding, evaluation of the above gives
\eqn\secondaryresult{
\sigma_m(X^{-1}) = \Gamma(1-m) X^{-m-1} = \Gamma(1-m) T_{-m}}
Now we put this in a correlator with three other tachyons, all
of positive chirality:
\eqn\corrsecond{
\eqalign{
\l \sigma_m(T_0) T_{k_2} T_{k_3} T_{k_4}\r &=
\Gamma(1-m)\l T_{-m}T_{k_2} T_{k_3} T_{k_4}\r\cr
&= (1-m)\Gamma(1-m) = \Gamma(2-m)\cr}}
where we have used the tachyon 4-point function formula
Eq.\fourfinal, in the appropriate kinematic region.

Next, we consider the gravitational recursion relation. A
general derivation of the relations in the present model is
beyond the scope of this work, but we will use the principle of
continuation from the $k>0$ case as a guide. In that case, on
the sphere we have\wittenphase\
\eqn\recurone{
\eqalign{
\l \prod_{i=1}^4\sigma_{m_i}(\phi_{r_i})\r
=& \sum_r \l \sigma_{m_1-1}(\phi_{r_1})
\sigma_{m_2}(\phi_{r_2})
\phi_r\r \eta^{rs} \l \phi_s \sigma_{m_3}(\phi_{r_3})
\sigma_{m_4}(\phi_{r_4})\r\cr
&+ \sum_r \l \sigma_{m_1-1}(\phi_{r_1})\phi_r\r \eta^{rs} \l \phi_s
\sigma_{m_2}(\phi_{r_2})\sigma_{m_3}(\phi_{r_3})
\sigma_{m_4}(\phi_{r_4})\r\cr}}
(we have removed factors of $m_1$ on the RHS since these are
absorbed by an $m!$ discrepancy between the norms used in
Refs.\losev\ and \wittenphase). Now let us specialise to the
case where $m_2,m_3,m_4=0$, so the last three fields are
primaries. In that case, it follows (for $k>0$) that the second
term on the RHS above does not contribute, by U$(1)$ charge
conservation, and the fact that primaries and secondaries in
$k>0$ models cannot carry the same charge.

Thus, we first drop this term and then take over the recursion
relation for our case. Then we have
\eqn\recurtwo{
\l \sigma_m(T_0) T_{k_2}T_{k_3}T_{k_4}\r =
\sum_k \l \sigma_{m-1}(T_0) T_{k_2}T_{k}\r \eta^{kk'}
\l T_{k'}T_{k_3}T_{k_4}\r}
On the RHS we find that the answer is $\Gamma(2-m)$ (from
Eq.\secondaryresult) times two momentum-conserving 3-point functions,
which are equal to 1. Thus the answer is again $\Gamma(2-m)$, in
agreement with Eq.\corrsecond.

Let us now re-introduce the cosmological constant, to provide an
additional confirmation of the scenario that we have developed.
We repeat the calculations in this section, but starting with the
scaled superpotential $\mu X^{-1}$. From Eq.\secondary\ the
result is that the negative-chirality tachyons are scaled by a
power of $\mu$ (while the positive ones are of course
unaffected). The result is:
\eqn\scale{
\eqalign{
T_k &\rightarrow T_k\qquad (k>0)\cr
T_k &\rightarrow (-\mu)^{|k|} T_k\qquad (k<0)\cr}}
With these scalings, we can repeat our computation of the
perturbed structure constants, Eq.\fourptpert. Suppose we take
$k_4>0$, $k_1,k_2,k_3<0$ in that equation, and scale the
superpotential and negative chirality tachyons as described
above. The result, after differentiating in $t_4$, is
proportional to $\mu^{-k_1-k_2-k_3-2}$. But this is {\it
precisely} the cosmological constant dependence of the tachyon
4-point function! Indeed, it is an easy exercise to check that in
general the above scaling gives rise to the following
$\mu$-dependence:
\eqn\cosmdep{
\l T_{k_1}\cdots T_{k_N}\r \sim \mu^{2-N + \half\sum_i |k_i|}}
which is precisely what is expected from matrix models or Liouville
theory. This confirms not only that the superpotential should be
thought of as the cosmological operator, but also that the
construction of negative-chirality tachyons as secondaries is
consistent.

The above observations provide a very interesting way to relate the LG
theory being discussed here with the Kazama-Suzuki model of
Ref.\mukhivafa. Taking $\mu\rightarrow 0$ simply sets the
superpotential to 0, leaving behind what we might call the ``free'' LG
model. (This is of course a singular change, as the resulting matter
theory has ${\hat c}=1$). Now, the result is {\it precisely} the
theory obtained in Sec.(3.4) of Ref.\mukhivafa, by starting with the
Kazama-Suzuki coset and describing the SL$(2)$ currents by a dual
Wakimoto representation. This provides a direct link between the KS
and LG pictures of the topological theory.

\beginsection 7. More on Gravitational Descendants

It is clear from the preceding section that the recursion relations
obeyed by the negative-momentum tachyons, treated as gravitational
descendants, are {\it not} the ones that would be obtained by just
taking over the results of Ref.\wittenphase,\kkli\ for the case of
unitary topological matter coupled to topological gravity. In that
context, one has to sum over all degenerations of the Riemann surface
and at the point of degeneration, one has to sum over a complete set
of matter primaries flowing through the pinch. This has no obvious
analogue in the present case, simply because the gravitational
primaries (positive tachyons) are dual to gravitational descendants
(negative tachyons) in our model. The main point of difference seems
to arise from the fact that our matter system is ``nonunitary'' as a
superconformal theory before twisting.

One may conjecture that the right gravitational recursion relations
are simply those of Ref.\wittenphase\ in which all the tachyons, both
primary and secondary (in dual pairs) flow through the pinch. But this
relation does not then reproduce the right tachyon $N$-point functions
that we have already derived in a previous section. Instead, we use
the result of the previous section to conjecture a modified
gravitational recursion relation. We will see that our conjecture is
powerful enough to give not only the $N$-point functions on
the sphere, but also the correct 2-point function on the torus.

{}From Eq.\recurtwo\ we see that out of two possible degenerations of
the sphere with four punctures, the one which contributes is the one
with the {\it minimum} number of fields in the ``right-hand'' branch
(the branch on which the gravitational descendant itself does not
lie). Thus we are led to postulate the following general relation for
an $N$-point function with a single gravitational secondary:
\eqn\recursphere{
\l \sigma_m(T_0) T_{k_2}\cdots T_{k_N}\r_0 =
\sum_k \l \sigma_{m-1}(T_0) T_{k_2}\cdots T_{k_{N-2}} T_{k}\r_0 \eta^{kk'}
\l T_{k'}T_{k_{N-1}}T_{k_N}\r_0}
The subscript on the correlators refers to the genus (0 in this case),
this has been displayed explicitly as we will shortly turn to the case
of the torus.

To check that this relation agrees with the known tachyon correlators
is a simple exercise. Using Eq.\nptfn, the LHS of the above equation
is
\eqn\recurlhs{
\eqalign{
\l \sigma_m(T_0) T_{k_2}\cdots T_{k_N}\r_0 &=
\Gamma(1-m)\l T_{-m} T_{k_2}\cdots T_{k_N}\r_0\cr
&=\Gamma(1-m)\prod_{i=1}^{N-3}(-m+i)\cr
&=\Gamma(N-2-m)\cr}}
while momentum conservation and Eq.\nptfn\ give for the RHS
\eqn\recurrhs{
\eqalign{
\sum_k \l \sigma_{m-1}(T_0) T_{k_2}&\cdots T_{k_{N-2}} T_{k}\r_0 \eta^{kk'}
\l T_{k'}T_{k_{N-1}}T_{k_N}\r_0 \cr
&=\l \sigma_{m-1}(T_0) T_{k_2}\cdots T_{k_{N-2}} T_{k_{N-1}+k_N-1}\r_0
\l T_{-k_{N-1}-k_N}T_{k_{N-1}}T_{k_N}\r_0\cr
&= \Gamma(2-m)\l T_{1-m} T_{k_2}\cdots T_{k_{N-2}}T_{k_{N-1}+k_N-1}\r_0\cr
&= \Gamma(2-m)\prod_{i=1}^{N-4}(1-m+i)\cr
&= \Gamma(N-2-m)\cr}}
Thus we see that the postulated recursion relation gives the correct
answer for $N$-point functions on the sphere as long as just one
gravitational descendant is present.

The above hypothesis for the sphere recursion relation, can now be
applied to the torus. Again we postulate that of the various ways in
which the torus can degenerate into a sphere times a torus, the
minimum number of fields (in this case, a single one) lie in the
``right branch'' (the torus). In addition, there is the usual term
corresponding to pinching a nontrivial cycle of the torus. Again, we
restrict ourselves to the case where only one of the inserted fields
is a gravitational secondary (negative tachyon). Thus we have
\eqn\recurtorus{
\eqalign{
\l \sigma_m(T_0) T_{k_2}\cdots T_{k_N}\r_1 =& {1\over 24}
\sum_{k} \l \sigma_{m-1}(T_0)T_{k_2}\cdots T_{k_N} T_k T_{-k-1}\r_0\cr
&+ \sum_k \l \sigma_{m-1}(T_0) T_{k_2}\cdots T_{k_N} T_k \r_0
\l T_{-k-1} \r_1\cr}}
Note that while the first term involves an infinite sum over all
positive and negative momenta, the second term has just a single
contribution from $k=-1$ by momentum conservation.

We will check this for the two-point functions. On the LHS, we have
\eqn\recurtoruslhs{
\eqalign{
\l \sigma_m(T_0) T_{k_2} \r_1 &= \Gamma(1-m)\l T_{-m} T_{k_2}\r_1\cr
&= \Gamma(1-m)\l T_{-m} T_m  \r_1\cr}}
where we have $k_2=m$ by momentum conservation, and the delta-function
has been suppressed. On the RHS we have
\eqn\recurtorusrhs{
\eqalign{
{1\over 24}&\sum_k \l \sigma_{m-1}(T_0)T_m T_k T_{-k-1}\r_0
+ \l \sigma_{m-1}(T_0) T_m T_{-1}\r_0 \l T_0 \r_1\cr
&= \Gamma(2-m)\left( {1\over 24} \sum_k \l T_{1-m} T_m T_k T_{-k-1}\r_0
+ \l T_0 \r_1 \right)\cr}}
It follows that our recursion relation in this case is equivalent to
\eqn\recurtwopt{
\l T_{-m} T_m \r_1 = (1-m) \left( {1\over 24}\sum_k \l T_{1-m}
T_m T_k T_{-k-1}\r_0 + \l T_0 \r_1 \right)}
The right-hand side is easily evaluated by dividing the sum over $k$
into regions corresponding to the distinct kinematic configurations
in the sphere four-point function. We have to isolate a divergent term
of the form $\sum_{k=1}^\infty k = \beta$, and the result is
\eqn\twopt{
\l T_{-m} T_m \r_1 = -{1\over 24} (1-m) \left( m^2 -m + \beta - 24\l
T_0 \r_1 \right)}

Although we do not know how to give a meaningful value to the infinite
constant $\beta$ in a way that is uniquely dictated by the physics of
this model, it can be fixed purely by self-consistency. To do this,
recall that the torus partition function is proportional to ${\rm
log}\mu$. Setting $m=0$ in the above equation and rewriting the
correlators of cosmological operators as derivatives in $\mu$ of the
partition function, we find that for consistency, we must have
$\beta=0$.

Finally we insert the value $\l T_0 \r_1 = {1\over 12}$ in
Eq.\twopt\ above to get
\eqn\twoptfinal{
\l T_{-m} T_m \r_1 = -{1\over 24} (1-m) \left(m^2 -m - 2 \right)}
which is precisely the matrix-model result\ref\kleblowe{\KLEBLOWE}!

Note that several correlators in the Landau-Ginzburg theory we have
been studying tend to have zeroes when some momentum $k$ is equal to
1. This can in each case be traced to the following fact: the tachyon
with unit momentum is represented in the LG theory by the identity
operator. When treated as a perturbation of the superpotential, this
clearly causes no change in $W'$, which accounts for the zeroes at
momenta $k=1$. Of course, this holds only for those correlators where
the kinematics is such that the only dependence on the given momentum
comes from perturbing the superpotential. Contact terms give rise to a
different dependence.

It should be clear that the gravitational recursion relations that we
have been using above are not quite the standard ones for topological
gravity --- however, it is important that they involve a {\it subset}
of the terms which appear in the standard case, and not any additional
terms. We have conjectured the truncation to this subset of terms and
shown that this consistently leads to correct results, but it would be
worth trying to find a proof of these recursion relations. In any
case, the discussion here applies only to the case of one
gravitational secondary and the remaining fields primary, while the
general case remains open.

We believe that with better understanding, this topological formulation
of $c=1$ string theory could be powerful enough to reproduce all
higher-genus correlators, thus rivalling the highly successful matrix
models.

\beginsection 8. Discussion and Conclusions

We have provided ample evidence that LG theory with superpotential
$X^{-1}$ describes the $c=1$ string where the matter field is
compactified on a circle with the self-dual radius. The $N$-point
functions of tachyons on the sphere and the two-point function on the
torus have been computed above, but in principle it should not be
difficult to go further. The main obstacle to that is a complete
understanding of the nature of gravitational recursion relations. The
conjecture which we have presented and tested should prove useful in
such an investigation.

The various discrete states of $c=1$ string theory have not yet been
clearly understood in the topological approaches. In Ref.\mukhivafa,
although they are clearly identified in the CFT description (which is
in terms of the Hilbert space at $\mu=0$), they are not
straightforward to identify in the Lagrangian description of the same
coset model (which, however, is automatically at nonzero $\mu$). The
same is true in the LG approach.

The obvious candidates for the states of ghost number 1 ($Y^+_{s,n}$
in the notation of Ref.\witzwie) come from the following
identification with gravitational descendants:
\eqn\disc{
Y^+_{s,n} \sim \sigma_m(T_k) \quad\quad (s=\half(m+k),~n=\half(m-k))}
with $k>0$. This has many appealing features: for the special cases
$k=0$ or $m=0$ it reduces to the identifications we have already
demonstrated above. Matter momentum conservation on the LHS gives the
same relation as U$(1)$ charge conservation on the RHS. Further
confirmation will require the computation of correlation functions,
for which we again need a better understanding of the gravitational
recursion relations. Also, there is no sign of winding modes, as the
above correspondence accounts only for the left-right symmetric
(momentum) modes.

Let us comment on the mysterious way in which the topological theory
discussed here fails to show explicit $\ZZ_2$ invariance, which in
$c=1$ string theory is simply sending the $c=1$ free scalar field to
minus itself (this is variously interpreted as time-reversal or
parity)\foot{We are grateful to E. Witten for stressing this point in
the context of the coset description of Ref.\mukhivafa, which
evidently has the same feature.}. In the LG theory discussed here, such
a symmetry would have to take $X^{k-1}$ to $X^{-k-1}$ which is clearly
not obtainable by any transformation on the $X$ superfield.  This
seems to originate in the fact that there is a shift of 1 unit between
the tachyon momentum and the topological U$(1)$ charge. One suggestive
observation is the following: a generic perturbation of the LG theory,
which can be thought of as the string field $\Psi(X)$, is given by
\eqn\sfield{
\Psi(X)=\sum_k t_k X^{k-1}}
If we interpret $X$ as a complex variable, this is precisely the mode
expansion of a spin-1 current in a conformal field theory. If
$\Psi(X)$ were really treated as a spin-1 conformal field, then the
inversion $X\rightarrow X^{-1}$ would effect the change $k\rightarrow
-k$ precisely because a spin-1 field picks up the appropriate Jacobian
when the transformation is carried out. Moreover, transforming the
variable to the cylinder by $X=e^Z$ would eliminate the `shift' by 1
unit, and we would have perfect $\ZZ_2$ symmetry on the cylinder.
This suggests that we need to understand better the `target-space'
properties of our theory, and the right variables in which to describe
it. It is noteworthy that a target-space conformal dimension of 1 for
the string field corresponding to the tachyons of $c=1$ string theory
has already been suggested by Witten and Zwiebach\witzwie. In their
work, this comes about by studying the transformation properties under
the target-space Virasoro algebra which arises as a subalgebra of
$W_\infty$.

The fact that the LG theory discussed in this paper has ${\hat c}=3$
strongly suggests that it is even more closely linked to Calabi-Yau
(CY) sigma-models than the conventional LG models, where one needs to
take many copies to get the correct central charge\ref\cy{\CY}.
Because the weight of the basic superfield $X$ is negative, it seems
likely that the related manifold will actually be a non-compact
analogue of a CY hypersurface, in a weighted projective space with
some negative weights.

The genus-$g$ partition function of this theory is expected to be the
Euler characteristic of the moduli space of genus-$g$ Riemann
surfaces\ref\penner{\PENNER}\wittenmatrix\mukhivafa. If this could be
computed directly from the present LG formulation, it would give
an independent derivation of the fact that the Euler characteristic of
moduli space is proportional to the Bernoulli numbers. Even more
interesting, the generating function for tachyon correlators in
genus-$g$ for the self-dual radius theory is known\ref\dijkmp{\DIJKMP}
in the form of a Kontsevich-like matrix integral. It would be
wonderful to derive this elegant integral representation directly from
topological arguments in Landau-Ginzburg theory.

\beginsection Acknowledgements

We are grateful to Sanjay Jain, T. Jayaraman, Varghese John and Ashoke
Sen for many helpful discussions and comments, and to Sourendu Gupta
for a useful suggestion. One of us (S.M.) is grateful for the warm
hospitality of the following institutes, where part of this work was
done: Center for Theoretical Studies, Bangalore; Institute of
Mathematical Sciences, Madras; and Institute of Physics, Bhubaneswar.

\listrefs
\bye